\def\be{\begin{equation}}
\def\ee{\end{equation}}
\def\ber{\begin{eqnarray}}
\def\eer{\end{eqnarray}}
\def\bers{\begin{eqnarray*}}
\def\eers{\end{eqnarray*}}

\documentclass[aps, prl, amsmath,amssymb, reprint, showpacs,superscriptaddress,groupedaddress, nofootinbib ]{revtex4-1}
\usepackage{dcolumn}   
\usepackage{amsmath}
\usepackage{graphicx}    
\usepackage{comment}
\usepackage{subfigure}
\usepackage{color}
\usepackage{multirow}
\usepackage{nicefrac}
\usepackage{soul}

\usepackage{ifthen}
\newboolean{includefigs}
\setboolean{includefigs}{true}      
\newboolean{includetext}
\setboolean{includetext}{true}     
\newcommand{\condcomment}[2]{\ifthenelse{#1}{#2}{}}
\begin{document}

\title{Bismuth based Half Heusler Alloys with giant thermoelectric figure of merit}
\author{Vikram,$^\dagger$ Jiban Kangsabanik,$^\dagger$ Enamullah$^\dagger$ and Aftab Alam}
\email{{\it aftab@phy.iitb.ac.in}}
\affiliation{Department of Physics, Indian Institute of Technology, Bombay, Powai, Mumbai 400 076, India}

\let\thefootnote\relax\footnote{$^\dagger$ These three authors have contributed equally to this work}

\begin{abstract}
Half Heusler (HH) thermoelectric alloys provide a wide platform to choose materials with non-toxic
and earth abundant elements. This article presents an {\it ab-initio} theoretical evaluation of electrical and thermal transport properties of three Bismuth-based most promising thermoelectric alloys, selected out of 54 stable HH compounds. These are brand new compounds which are recently proposed to be stable (Nature Chem. {\bf 7}, 308 (2015)) and may have interesting properties. The calculated band structure of the three compounds, namely HfRhBi, ZrIrBi and ZrRhBi, served as a hint for their promising thermoelectric properties. To gain confidence on the theoretical predictions of these unreported systems, we first checked our calculated results for a well studied similar compound, ZrNiSn, and showed reasonable agreement with the measured ones. HfRhBi and ZrIrBi turn out to be narrow band gap while ZrRhBi is a moderate band gap semiconductor. A detailed study of the carrier concentration and temperature dependance of the Seebeck coefficient (S), Power factor (S$^2 \sigma$), lattice ($\kappa_L$) and electronic ($\kappa_e$) thermal conductivity and hence the figure of merit ($Z$T) is carried out. In contrast to most promising known thermoelectric materials, we found high power factor for these materials (highest S$^2 \sigma\sim$17.36 mWm$^{-1}$K$^{-2}$ for p-type ZrIrBi). All the three systems (specially p-type) show high figure of merit, with ZT value as high as 0.45 for ideal crystal. Maximum ZT and the corresponding optimal n- and p-type doping concentrations ($n_c$) are calculated for all the three compounds, which shall certainly pave guidance to future experimental work.

\end{abstract}

\maketitle
\section{Introduction}
{\par}   
The ever increasing demand of energy rests on a concerted effort to broaden our energy resources (solar, fuel, wind, etc.) and at the same time reduce our energy consumption. As a matter of fact, a large fraction of energy consumed  throughout the globe is lost as heat, and conversion of even a small fraction of it to other form of usable energy can significantly impact the globe energy requirement. Significant efforts are underway to convert these waste heat into electricity using thermoelectrics. In addition, advances in thermoelectrics lead to the development of multistage Peltier coolers which can be used to achieve efficient cooling/heating.

The fundamental requirement to obtain a promising thermoelectric material is to optimize a variety of conflicting properties. The conversion efficiency of thermoelectric (TE) materials depends
upon the transport coefficients of the constituent material through
a dimension-less figure of merit, ZT=(S$^{2}\sigma$T)/($\kappa_{e}$+$\kappa_{L})$.
Here, S, $\sigma$ and T are the Seebeck coefficient, electrical conductivity and temperature respectively.  $\kappa_{e}$ ($\kappa_{L}$) are the thermal conductivity due to electrons (phonons).

The value of ZT measures how efficient the material is for both TE
heating/cooling applications, higher the value higher the efficiency. The main challenge
among the researchers is how to enhance the value of ZT since the transport
parameters S, $\sigma$ and $\kappa$ ( $\kappa_{e}$ and $\kappa_{L}$)
are inter-related to each other. The electronic thermal conductivity and electrical conductivity is related to each other
by the Wiedemann-Franz law, $\kappa_{e}$ = L$\sigma$T,
where L is the Lorentz number. This relation clearly shows that increasing
electrical conductivity also increases electronic thermal conductivity
hence optimization of ZT turns out to be challenging.
In earlier days, metals were used for TE applications because of very
good electrical conductivity but their ZT values were low since metals possess large thermal conductivity with significant contribution from the lattice part ($\kappa_{L}$). Hence minimizing lattice thermal conductivity
is itself a challenge.
Most popular TE materials are based upon Bi$_{2}$Te$_{3}$ and Si$_{1-x}$Ge$_{x}$ having ZT value
above 1 at their optimal temperature.\cite{key-1}$^{,}$\cite{key-2} Other bulk materials,
such as complex chalcogenides,\cite{key-3}$^{,}$\cite{key-4} skutterudites\cite{key-5}$^{,}$\cite{key-6} and
quasi-crystals\cite{key-7} are potential candidates for TE applications.

Heusler alloys have been known to mankind since the beginning of the
20\textsuperscript{th} century. They are known to have a very rich
magnetic behaviour which make them a promising candidate for spintronic
applications.\cite{key-8} Good electrical and mechanical properties, thermal stability
and easy tunability of band gap (0-4 eV) by only changing the chemical
composition makes them potential candidates for solar cell applications
and topological insulators\cite{key-8}. But in the last decade, ternary
half-Heusler alloys (HHAs) have emerged as promising candidates
for TE applications also due to small value of band gap, high
value of Seebeck coefficient (S) (upto several hundreds $\mu$VK$^{-1}$)
and its appealing transport properties.

A very big advantage of considering HHAs for thermoelectric application
is that they can be easily synthesized as 100\% dense samples\cite{key-9}
and the efficiency (i.e. ZT) can be enhanced by the application of
easy doping (either with n-type or p-type) or by inducing impurities
or defects. Another reason to consider HHAs
is the high value of their power factor (S$^{2}\sigma$) which
is one of the most important criteria for TE applications.

HHAs can serve as high as well as moderate temperature TE materials and have the potential to 
replace some of the state of art TE materials.\cite{key-19}$^{-}$\cite{key-29}
In the last few years, there has been extensive efforts to find new HHAs as potential TE materials. Until now, the most reliable HH compound found are the family of MNiSn (M=Ti, Zr, Hf) - related compounds with the maximum ZT values lying in range 0.7-1.5.\cite{APL79}$^,$ \cite{APL86}$^,$ \cite{acta.Mater.57} However, this (ZT)$_{max}$ is achieved by playing with various types of doping (e.g. mixing M-elements with each other or doping Sb with Sn etc.), nanostructuring etc. The maximum ZT obtained for ideal crystal of ZrNiSn (no doping), however, remains $0.2-0.3$.\cite{APL79} Among others, NbCoSn (ZT$\sim$ 0.15),\cite{APL92} TiCoSb (ZT$\sim$ 0.015),\cite{JAP384} etc. emerge out to be good TE HH compounds. The main reason for not getting high value of ZT in ideal crystals of HHAs is the high values of thermal conductivity and relatively lower values of power factor. For example, the experimentally determined maximum power factor for ZrNiSn-based alloy \cite{APL79} is 3.4 mWm$^{-1}$K$^{-2}$ at 750 K, and for TiCoSb-based alloy\cite{JAP102}  is 2.3 mWm$^{-1}$K$^{-2}$ at 850 K. Although, these power factors are relatively higher than those of most current TE materials, but not as high as to enhance the figure of merit to a great extent. On the other hand, ultra low thermal conductivity has been found recently in YNiBi system, \cite{JAP117} but with a very low ZT.

\begin{figure}[t]
\begin{centering}
\includegraphics[scale=0.35, trim={0 100 0 100},clip]{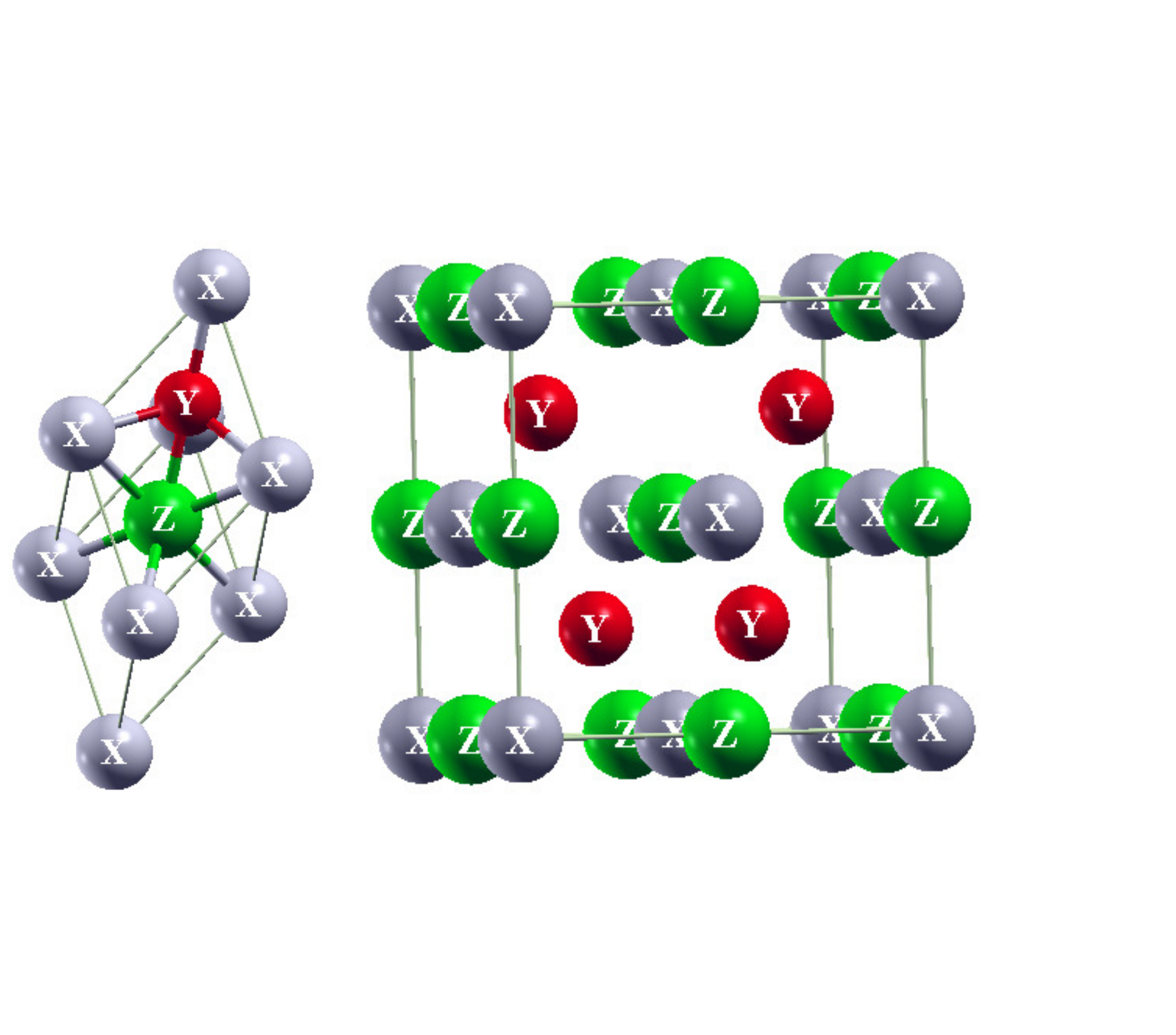}
\par\end{centering}
\caption{A 3-atom primitive cell (left) and a cubic unit cell (right) of HHAs
XYZ, where X represents either Hf or Zr, Y as either
Rh or Ir; and Z as Bi.}
\label{cell}
\end{figure}

Although, there exists several ideal HH crystals with relatively high ZT values, however the hunt for more such alloys is still ongoing in order to find better TE materials. In fact, if one look at the Inorganic Crystal Structural Database (ICSD) \cite{ICSD} or Pearsons hand book \cite{Pearson+handbook} there exists hundreds of HH alloys. More recently, a huge database\cite{nat.chem} is created for such alloys with the total valance electron count (VEC) in their primitive unit cell to be 18. Their main motivation was to report chemically plausible compounds that can have interesting properties but were neglected in the past. They mainly checked the stability and some basic electronic structure properties of these compounds using high throughput first principle calculations. Because HH compounds with VEC=18 have narrow band gaps and show promising TE properties, we have done a detailed band structure calculation\cite{unpublished} of all the stable compounds reported in Ref. [\citenum{nat.chem}]. Topology of these band structures helped to identify three compounds HfRhBi, ZrIrBi and ZrRhBi which turn out to be quite promising with the maximum ZT reaching upto $\sim$0.45.

The purpose of this paper is to systematically evaluate and present a detailed electronic structure, electrical as well as thermal transport properties of these three alloys. The key features of the present study which make it superior (with a higher predictability) compared to other theoretical studies existing in the literature are as follows,
\begin{enumerate}
\item Most current theoretical work on the topic of thermoelectricity are simply based on the density of states, formation of energy gaps, \cite{JAC296}$^,$ \cite{JPD39}$^,$ \cite{JPCM15} related chemical interaction between the d-states of transition metal,\cite{JAC296}$^,$ \cite{JPD39} compositional dependence of electronic structure, etc. In more recent articles people have looked at some aspects of electrical transport as well.\cite{AFM18}$^,$ \cite{JAP113} The thermal transport arising from the phonons, however, is mostly ignored. In this paper we have considered both electrical and thermal transport on the same footing so as to report the most reliable values of ZT.
\item In most previous studies, Seebeck coefficient (S) or electrical conductivity ($\sigma$) or the power factor (S$^2\sigma$) \cite{AFM18} or at best S$^2\sigma$/$\kappa_e$ is chosen as the parameter to dictate the thermoelectric performance or the optimal n-type or p-type doping levels for the concerned compound. However the ideal parameter should have been ZT as at low and moderate temperatures the lattice thermal conductivity ($\kappa_L$) plays an important role. Here, we have chosen ZT to be the comprehensive parameter to do the same and all the thermoelectric parameters reported in this paper corresponds to that optimal (ZT)$_{max}$.
\item A detailed phonon based analysis (phonon dispersion, phonon density of states), usually ignored in most literature, is performed here (see supplement \cite{supl}) to better understand the effect of lattice dynamics. 
\item Mechanical stability for these class of compounds has been seldom reported.\cite{M1}$^-$\cite{M4} Keeping the device fabrication and application in mind, we have calculated the elastic constants, Bulk modulus (B) and the Young's modulus (Y) to assert the robustness of these systems against various stress.
\end{enumerate} 
\section{Computational and theoretical Details}
{\par}

\begin{table}[t]
\begin{ruledtabular}
\begin{centering}
\begin{tabular}{|c|c|c|c|}
system & $a(\mathring{A})$ & $\Delta E_{g}(eV)$ & $\Delta E_{f}$$(eV/atom)$\\
\hline 
HfRhBi & 6.41 & 0.17 & -0.651\\
\hline 
ZrIrBi & 6.48 & 0.26 & -0.315\\
\hline 
ZrRhBi & 6.44 & 1.02 & -0.725\\
\end{tabular}
\par\end{centering}
\caption{Band gap $(\Delta E_{g})$ and formation energy $(\Delta E_{f})$
at theoretically relaxed lattice constant $(a)$.}
\label{form_energy}
\end{ruledtabular}
\end{table}

${\it {Ab}-}{\it {initio}}$ simulations are performed by using density
functional theory (DFT)\cite{key-38} implemented within Vienna ${\it {ab}-}$
${\it {initio}}$ simulation package (VASP)\cite{key-39}$^-$\cite{key-41} with a projected
augmented-wave basis.\cite{key-42} We used the Generalized Gradient
Approximation (GGA) with Perdew-Burke-Ernzerhof (PBE)\cite{key-43} scheme for the electronic exchange-correlation functional. A plane wave cut-off of 500 eV was used in
all the calculations. Tetrahedron method with Bl\"{o}chl corrections \cite{blochl_correc} were
used to calculate the Density of states (DOS). Electronic structure optimization was converged within the energy error $\sim$10$^{-5}$ eV.
Cell volume, shape and atomic positions for all the structures were
fully relaxed using conjugate gradient algorithm till the forces on
each atom falls below 0.01 eV/$\r{A}$. The Brillouin zone sampling was done
by using Automated $\Gamma$-centered K point mesh. A 20$\times$20$\times$20 K point
mesh was used to do the ionic relaxation as well as the self consistent
calculations.

The thermoelectric properties of suitable candidates were then calculated
using the BoltzWann module\cite{key-46} of Wannier90 package.\cite{key-47}
This uses Maximally Localised Wannier Function (MLWF) basis set\cite{key-48}$^,$\cite{key-49} by interpolating first principle
plane wave results to determine the Seebeck coefficient (S), electrical conductivity
($\sigma$) and the electronic part of the thermal conductivity
($\kappa_{e}$ ) using semi classical Boltzmann transport theory. The calculations were performed within a constant relaxation time approximation scheme($\tau$=10 fs). For transport calculations, we took 24 bands and 20 wannier functions for all the three systems. The frozen window was taken around the Fermi level (E$_F$) in such a way that it included all the states withing 2 eV range above and below the conduction and valence band edges respectively. 

The properties for n-type
and p-type materials were evaluated based on rigid band approximation.
According to it the addition of donor or acceptor atom does not affect
the band structure of the material, but instead causes the chemical
potential to shift. The chemical potential moves towards the conduction
bands as electron doping increase and goes towards the valance bands
as the hole doping increases. The output of the BoltzWann package is in terms of properties varying 
with respect to the chemical potential($\mu$). In order to study the transport properties
in terms of the carrier concentration ($n_c$) instead of the chemical potential, we use the equation $n_c=\intop D(E)f(E)dE$,
where $D(E)$ is the density of states and $f(E)$ is the Fermi-Dirac 
distribution function for electrons of energy E. At
a fixed temperature, the integral can be performed over entire energy
range at a given chemical potential to give the corresponding carrier
concentration value or vice versa. 

To calculate the lattice contribution to the thermal conductivity
($\kappa_{L}$) we use Boltzmann transport theory for phonons as
implemented in ShengBTE package.\cite{key-50} For this, we calculated the
2nd order inter-atomic force constants (IFCs) using PHONOPY package\cite{key-51}. The
3rd order IFCs are calculated using thirdorder.py module of ShengBTE package using
 small displacement method which generate atomic configuration
files with required displacements. Here we took 4$\times$4$\times$4 super-cell to
accurately calculate the force constants for both 2nd order and 3rd
order IFC calculations. Interactions up to fourth nearest neighbour
is taken into account for the calculation of 3rd order force constants. The
Born effective charges on each atom and the macroscopic dielectric
tensor is calculated using Density functional perturbation theory
on the primitive cell using 16$\times$16$\times$16 k-mesh.

HHAs have a general composition of XYZ, where X and Y are
low and high electronegative transition elements and Z is a main
group element. HHA crystallizes in MgAgAs type structure 
(space group F-43m (\#216) ), as shown in Figure \ref{cell}.
Our structural optimization suggest that the most stable structure is
the one with X at 4a(0,0,0), Y at 4c(0.25, 0.25, 0.25) and Z
at 4b(0.5, 0.5 0.5) wyckoff positions. For the TE analysis, specially
to calculate the lattice thermal conductivity of the system, we made
a super cell of 4$\times$4$\times$4 from a 3-atom primitive cell (shown
in Figure \ref{cell}).

Further computational details about the formation energies and the mechanical properties are given in the supplimentary material.\cite{supl}

\section{Results and Discussion}

\begin{table}[t]
\begin{ruledtabular}
\begin{centering}
\begin{tabular}{|c|c|c|c|c|c|}
system & C$_{11}$  & C$_{12}$ & C$_{44}$ & $B$ & $Y$\\
\hline 
HfRhBi & 198.29  & 92.29  & 47.95  & 127.62  & 133.79 \\
\hline 
ZrIrBi & 211.60  & 94.19  & 55.28  & 133.33  & 149.64 \\
\hline 
ZrRhBi & 194.14  & 86.96  & 50.13  & 122.69  & 136.37 \\
\end{tabular}
\par\end{centering}
\caption{Calculated elastic constants C$_{ij}$, Bulk modulus $(B)$
and Young's modulus $(Y)$ (all in GPa) at theoretically relaxed lattice constant.}
\label{mech_stability}
\end{ruledtabular}
\end{table}
The detailed theoretical analysis of chemical stability, electronic
structure, mechanical stability and TE properties for the three systems
HfRhBi, ZrIrBi and ZrRhBi are summarized in this section. Table \ref{form_energy}
presents the theoretically relaxed lattice parameters, the formation energies and
the direct band gap values obtained from the electronic structure calculations. 
Large negative values of formation energies
make these systems quite stable to external changes. The
low band gap (\ensuremath{\le}1 eV) make them potential candidates as TE
material.
\begin{figure}[t]
\begin{centering}
\includegraphics[scale=0.45]{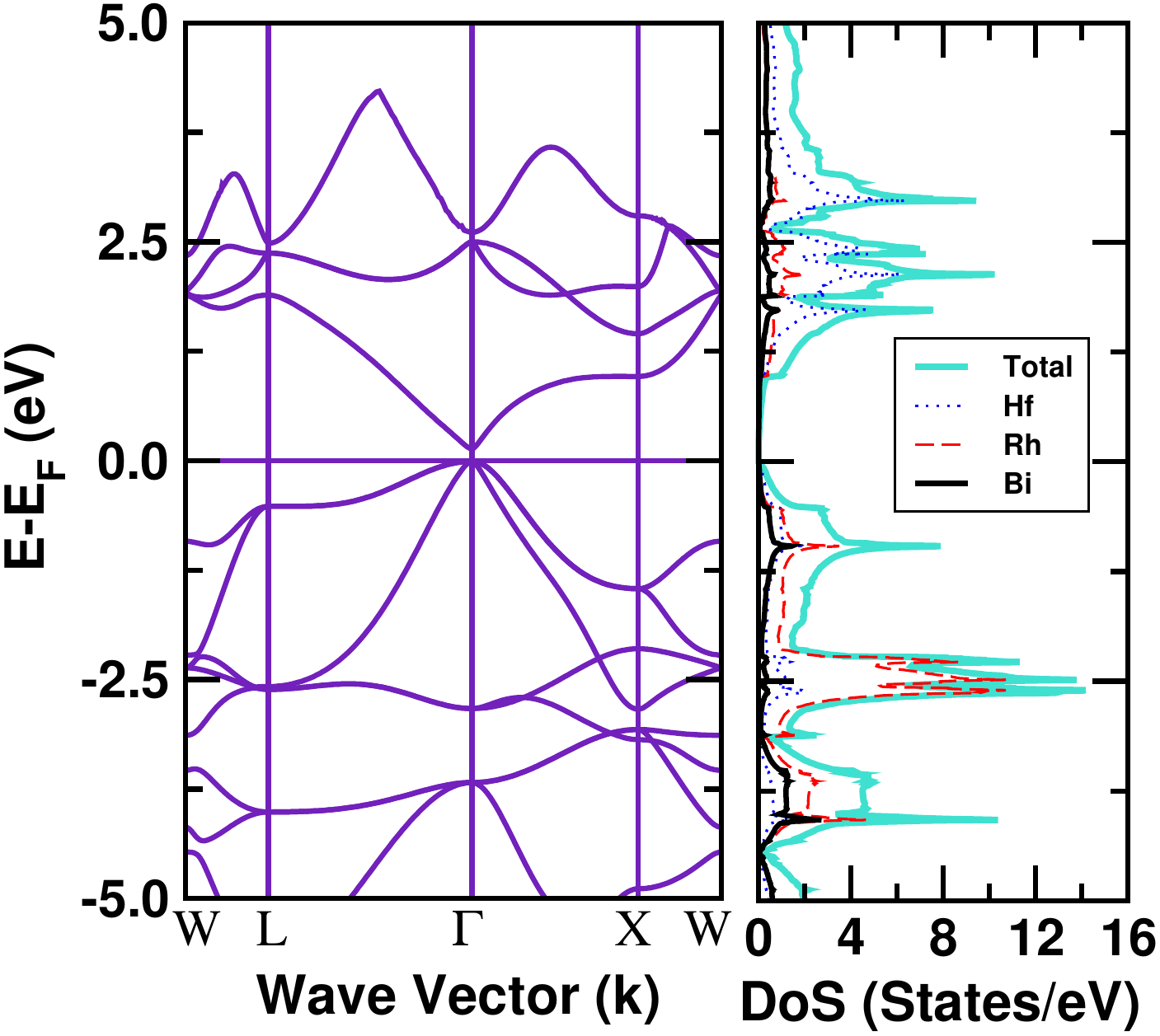}
\par\end{centering}
\caption{Band structure(left) and atom projected DOS(right) of
HfRhBi.}
\label{dos+bs_HfRhBi}
\end{figure}

\begin{figure*}[t!]
\begin{centering}
\includegraphics[width=7in,height=3in]{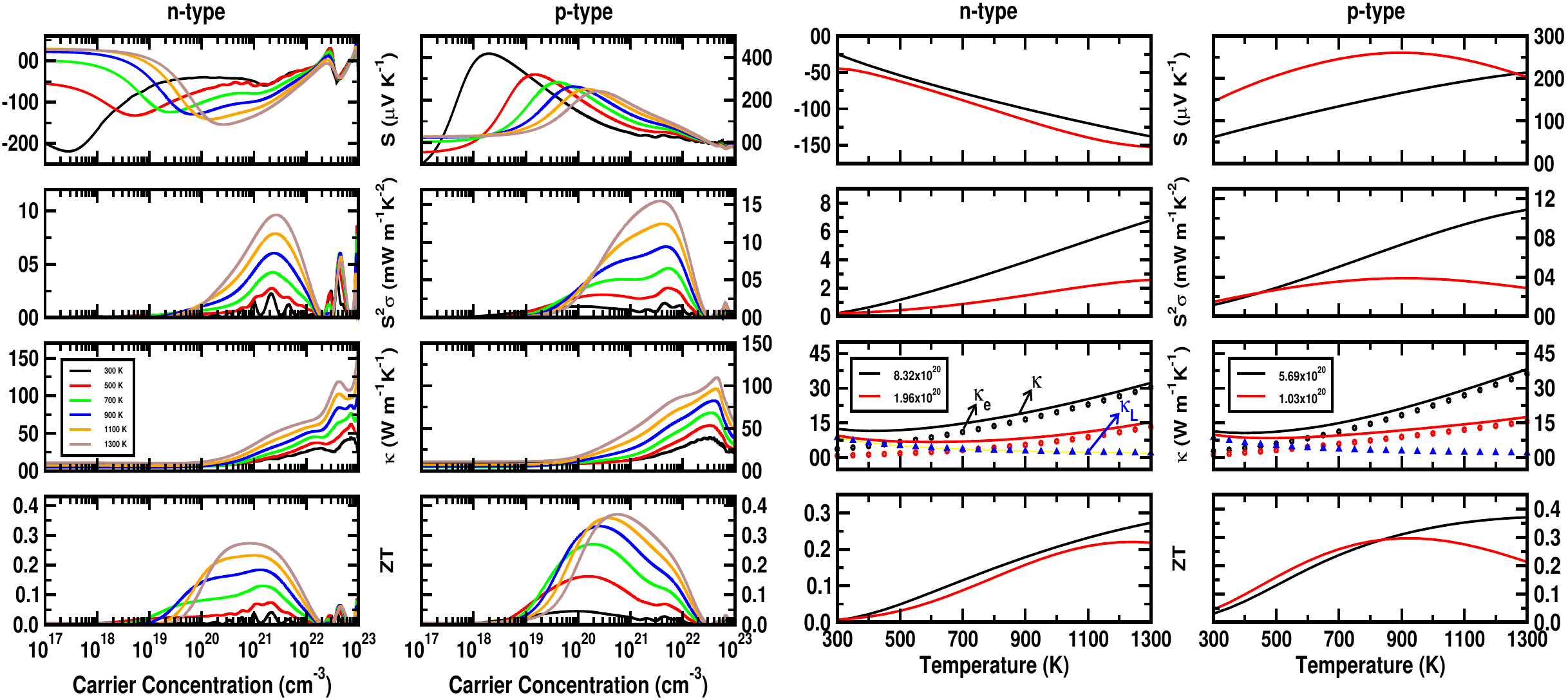}
\par\end{centering}
\caption{(Left) Carrier concentration dependence of Seebeck coefficients (S), power factor (S$^{2}\sigma$), thermal conductivity [lattice ($\kappa_L$), electronic ($\kappa_e$) and total ($\kappa$)]
and figure of merit (ZT) for n and p-type HfRhBi alloy at various temperatures(T). (Right) Temperature dependence of the same quantities for n and p-type HfRhBi at two different optimal concentrations.}
\label{HfRhBi}
\end{figure*}

As discussed, since the mechanical stability of the material is an
important property for it to be used for device fabrication, and hence for thermoelectric application, we also
calculated the elastic constants for the three systems. The relevant parameters
required to check the mechanical stability of the material
as per the Born-Huang criteria (see supplement\cite{supl} for details) are listed in Table \ref{mech_stability}. These parameters satisfy all the criteria required for mechanical stability of all the three systems. The calculated values of
 elastic constants (B and Y)  are comparable to those for standard metals. This suggests
that they can be cut or drawn into different shapes without changing
the properties up to a fairly high point.

\section{H$\text{f}$R$\text{h}$B$\text{i}$}

{\it Electronic Structure :} Figure \ref{dos+bs_HfRhBi} shows the band structure and atom projected DOS for HfRhBi. It has a direct band gap of 0.173 $eV$ at $\Gamma$
point. This low value of the band gap aids easy excitation of thermally
excited electrons increasing the electrical conductivity. Flat bands 
(causing a sharp DOS) at/near the valence band maxima 
 near E$_F$ suggest a high value of effective mass 
for holes implying that this material can be used as a potentially good
p-type TE material.  Since the conduction band states are
very sharp near E$_F$, the n-type figure of merit is expected
 to be lesser than the p-type. High bulk modulus ($\approx$127.6 GPa), 
Young's modulus ($\approx$133.8 GPa) (see Table \ref{mech_stability}) along with the strong chemical stability suggests the material to remain robust in structure and electronic properties against pressure or other stress.

{\it Thermoelectric properties:} Figure \ref{HfRhBi} shows the Seebeck coefficient (S), Power factor (S$^2 \sigma$), thermal conductivity $\kappa$ (both lattice and electronic) and the figure of merit (ZT) as a function of carrier concentration $n_c$ (left) and temperature (T) (right) for both n-type and p-type HfRhBi. The $n_c$ dependence of all the quantities are shown at six different temperatures (T). Two optimal doping concentrations [one each for( ZT)$_{max}$ at 1300 K and 900 K]  are then chosen to plot the temperature dependence (right panel) for the same quantities. 

As evident from the figure, the Seebeck coefficient for p-type first increases with $n_c$, reaching a maximum and then decreases with further increase in $n_c$. The behavior for n-type, however, is completely different. In the concentration
range $\sim$ 10$^{20}$-10$^{22}$cm$^{-3}$, the electrical conductivity ($\sigma$) increases rapidly, and dominate the behavior of the power factor vs. n$_c$. This corresponds to an optimal carrier concentration, ($n_c$)$_{max}$, based on the maxima in the power factor (S$^{2}\sigma$) at a given temperature. However, looking at the variation of thermal conductivity ($\kappa$) vs. $n_c$, $\kappa$ shows a maximum at a relatively higher concentration than the optimal concentration, as estimated using S$^2 \sigma$. This changes the value of the optimal carrier concentration at which ZT achieves the maximum value. As such, the most common practice \cite{AFM18} for estimating the optimal concentration based on a maxima in the S$^2 \sigma$ is misleading. The variation of the thermal conductivity and hence its effect on ZT, can completely change the value of (n$_c$)$_{max}$. We have chosen the $n_c$ dependence of ZT (instead of S$^2 \sigma$) to obtain the optimal carrier concentrations, ($n_c$)$_{max}$. The
temperature dependence is plotted for one more carrier concentration
having ZT value $\sim$ 80\% of the (ZT)$_{max}$ for comparison sake (1.96$\times$ 10$^{20}$ cm$^{-3}$ for n-type and 1.03$\times$ 10$^{20}$ cm$^{-3}$ for p-type).
\begin{figure}[b]
\begin{centering}
\includegraphics[scale=0.45]{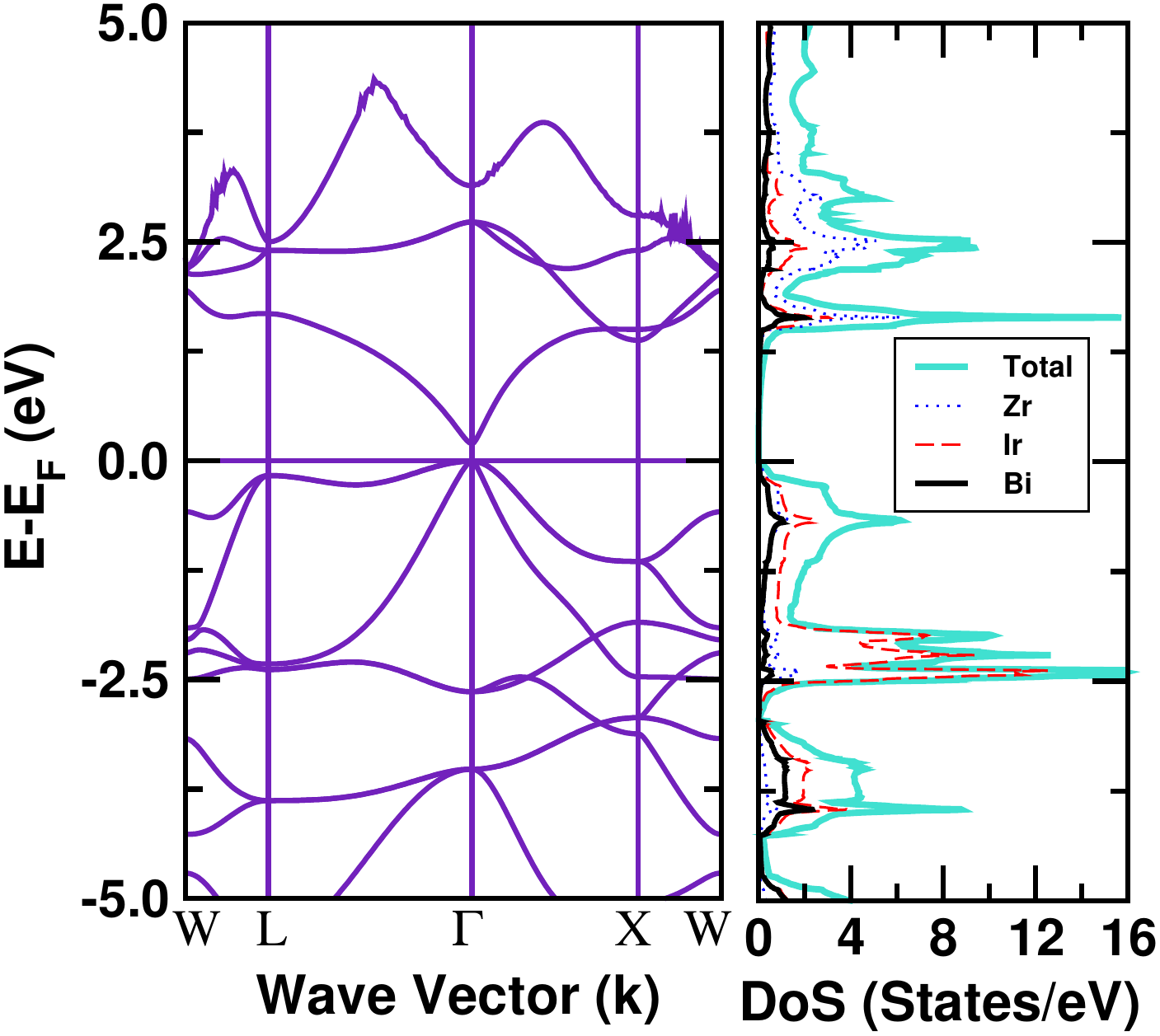}
\par\end{centering}
\caption{Band structure(left) and atom projected DOS(right) of
ZrIrBi.}
\label{dos+bs_ZrIrBi}
\end{figure}

\begin{figure*}[t]
\begin{centering}
\includegraphics[width=7in,height=3in]{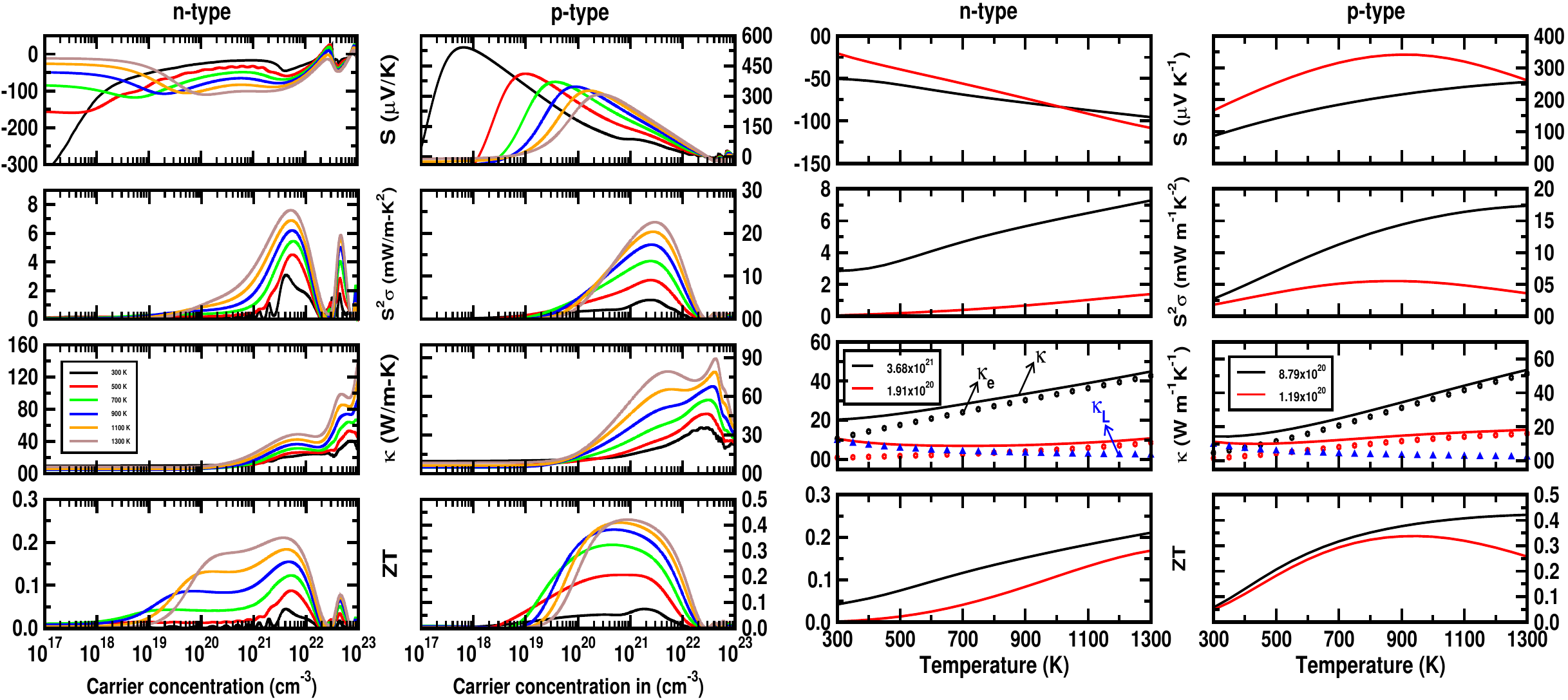}
\par\end{centering}
\caption{Same as Fig. \ref{HfRhBi}, but for ZrIrBi alloy.}
\label{ZrIrBi}
\end{figure*}

For p-type HfRhBi, the maximum power factor came out to be
$\approx$10.88 mWm$^{-1}$K$^{-2}$ at 1300 K. At such a high temperature, due to
thermally excited electron-hole pairs, the electronic part of thermal
conductivity ($\kappa_{e}$) becomes very large but at the same time
lattice thermal conductivity ($\kappa_{L}$) decreases drastically due to
decrease in the mean free path ($L_{ph}$) of the highly agitated phonons.
 The ZT value comes out to be 0.37. Even at smaller temperatures (800-1000 K),
power factor lies in the range  $\approx$6.2-8.4 mWm$^{-1}$K$^{-2}$ which is much larger
than the values obtained for most promising HH thermoelectric materials
e.g. $\approx$3.4 mWm$^{-1}$K$^{-2}$ for ZrNiSn, \cite{APL79}
$\approx$2.3 mWm$^{-1}$K$^{-2}$ for TiCoSb.\cite{JAP102}

For n-type HfRhBi at 1300 K,
the Seebeck coefficient is less than that of p-type.
This is due to a lower value of the effective mass in the n-region, as 
obvious from the band structure. The power factor and hence the ZT
is also smaller. The optimal value of ZT at $1300$ K is 0.27.
 
\begin{figure}[b]
\begin{centering}
\includegraphics[scale=0.45]{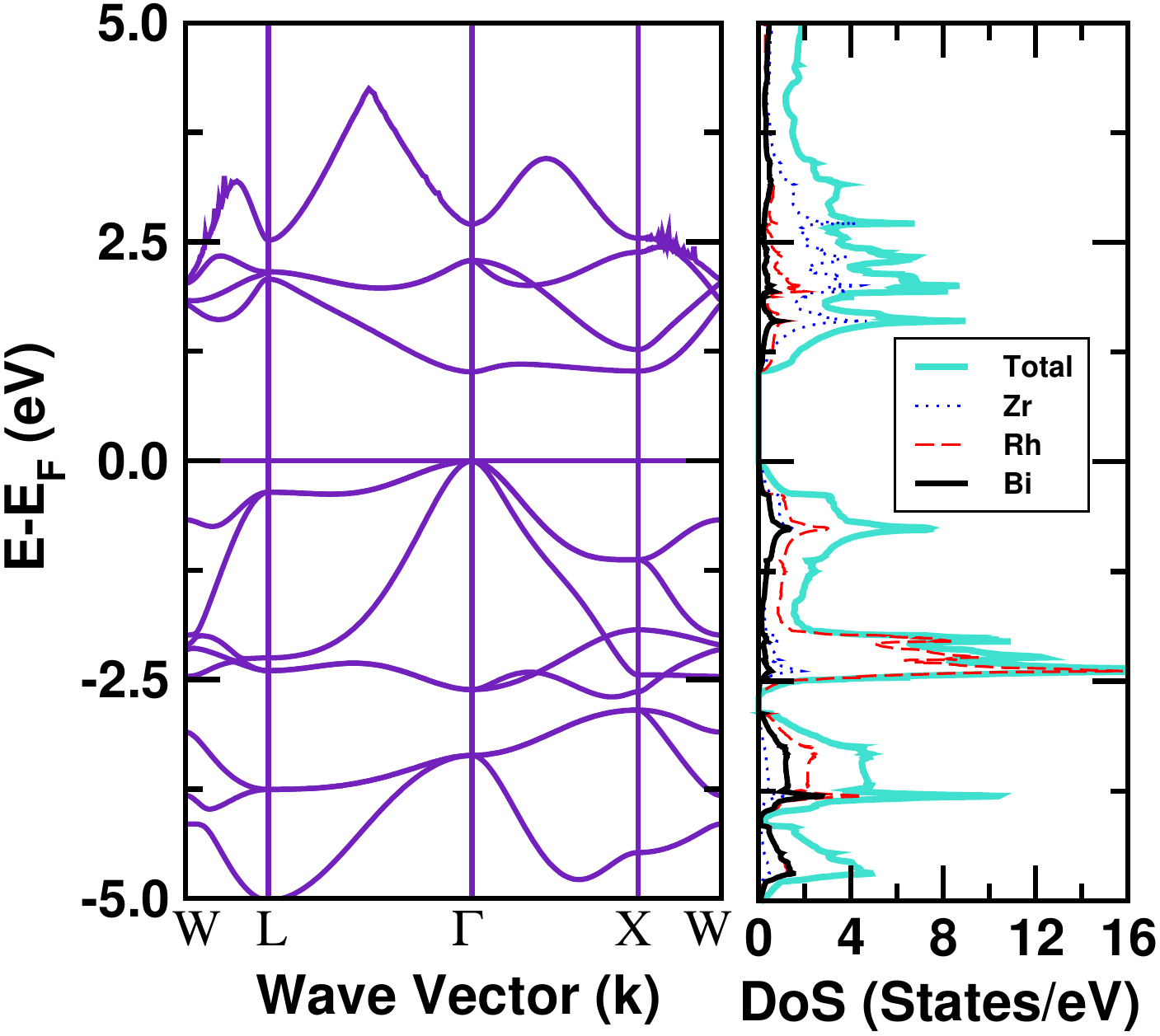}
\par\end{centering}
\caption{Band structure(left) and atom projected DOS(right) of
ZrRhBi.}
\label{dos+bs_ZrRhBi}
\end{figure}

\begin{figure*}[t]
\begin{centering}
\includegraphics[width=7in,height=3in]{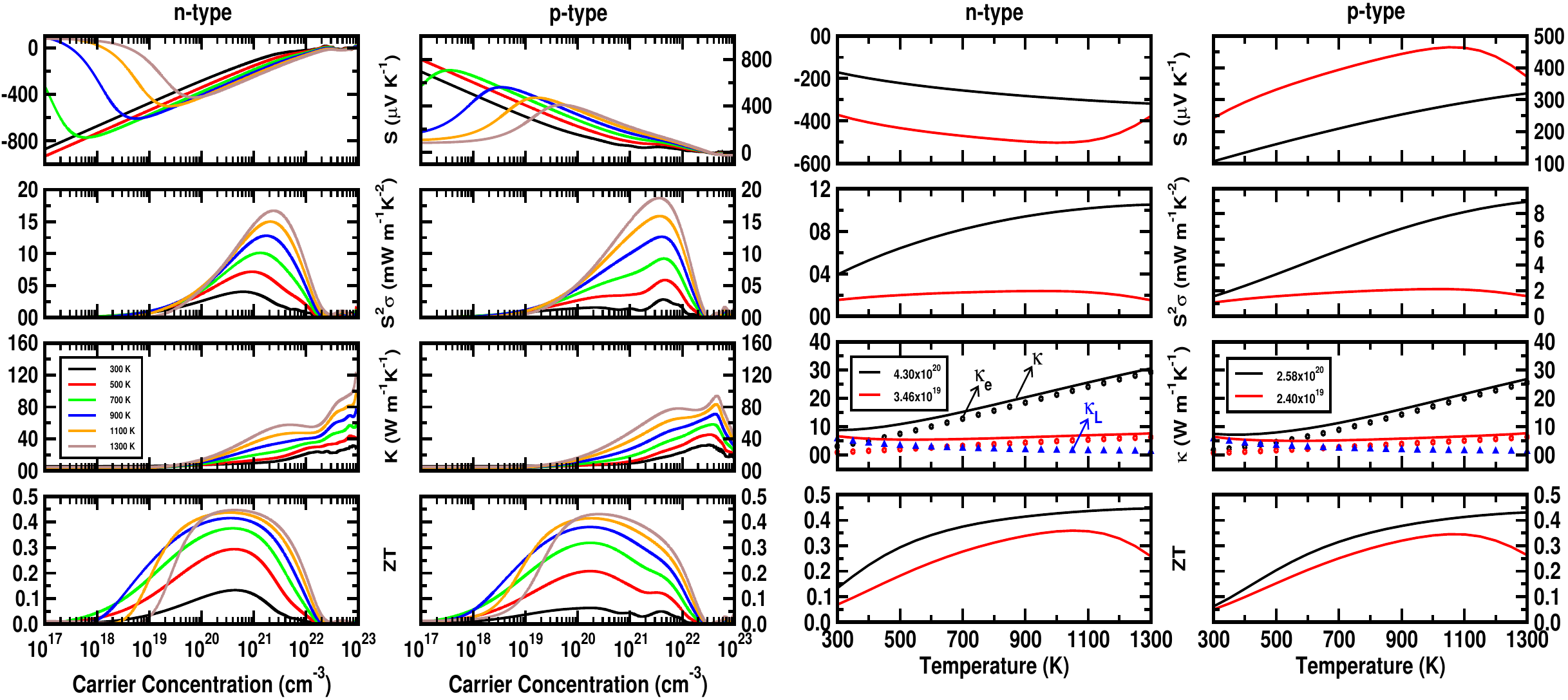}
\par\end{centering}
\caption{Same as Fig.\ref{HfRhBi}, but for ZrRhBi alloy.}
\label{ZrRhBi}
\end{figure*}

\begin{figure}[b]
\begin{centering}
\includegraphics[height=5cm, width=8.5cm]{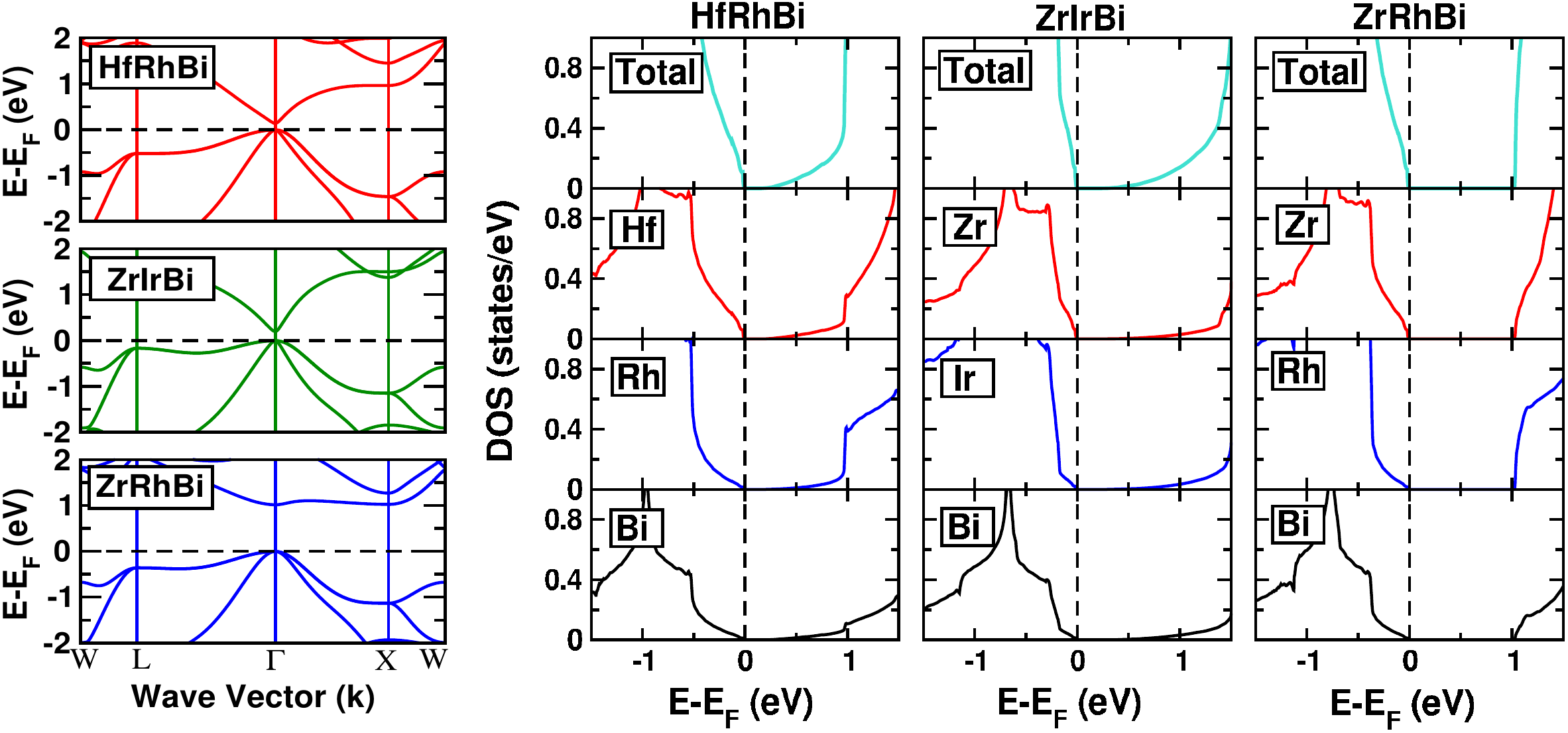}
\par\end{centering}
\caption{Closer look of (left) band structure and (right) total and atom projected DOS for HfRhBi, ZrIrBi and ZrRhBi near E$_F$ showing valence and conduction band edges.}
\label{dos+pdos}
\end{figure}

\section{Z$
\text{r}$I$\text{r}$B$\text{i}$}

{\it Electronic Structure :} Band structure and atom projected density of states for ZrIrBi are shown in Fig. \ref{dos+bs_ZrIrBi}. This system has similar band structure as HfRhBi near E$_F$ 
except for the fact that the top most valence band is more flat
and the conduction band edges are more steeper near $\Gamma$ point. Thus
one can expect the Seebeck coefficient
and power factor for ZrIrBi to be relatively higher than that of
HfRhBi for p-type system. Whereas the steeper conduction bands and
a very slow rise in the value of DOS above E$_F$ suggest it
to have lower S and hence ZT value than HfRhBi for n-type system. ZrIrBi has highest
mechanical stability out of the three systems with the bulk modulus $\approx$133.3 GPa
and Young's modulus $\approx$149.6 GPa (Table \ref{mech_stability}).

{\it Thermoelectric properties:} The variation of relevant quantities ($S$, $S^{2}\sigma$, $\kappa$
and $ZT$) with carrier concentration (left) and temperature (right) are shown in
Fig. \ref{ZrIrBi}. Similar to HfRhBi, ZrIrBi is a good p-type TE
material. Compared to HfRhBi, it has a large power factor and low lattice thermal conductivity ($\approx$2
Wm$^{-1}$K$^{-1}$) at intermediate as well as higher temperature range. 
At optimal carrier concentration ($n_c$)$_{max}$ = 8.79$\times$ 10$^{20}$cm$^{-3}$, it has a Seebeck Coefficient of 255.9 $\mu$VK$^{-1}$) and electrical conductivity of 26.5$\times$10$^4$ $\Omega^{-1}m^{-1}$ yielding the  power factor to be as large as $\approx$17.4 mWm$^{-1}$K$^{-2}$ (at T=1300 K). This gives a ZT value of 0.42. Notably, the ZT-value does not reduce much until 900 K (ZT$\approx$ 0.38). As such, p-type ZrIrBi can be used as a promising TE material even at relatively low T. 


For n-type system, one can notice a shallower increase in the
 DOS near conduction band edge arising out of steep bands near E$_F$ (see
Fig. \ref{dos+bs_ZrIrBi}),
thus making the effective mass relatively small and hence giving smaller value
of Seebeck coefficient (95.6 $\mu$VK$^{-1}$ at 1300 K). Since at high temperature,
the contribution of $\kappa_{L}$ to total thermal conductivity is very small, so ZT effectively depends on the S value ($ZT\approx S^{2}/L$, where $L$ is the Lorentz number).
Thus in spite of having a high $\sigma$ value (79.7$\times$10$^4$ $\Omega^{-1}m^{-1}$, which is almost double to that of HfRhBi), ZT (0.21) for ZrIrBi turn out to be smaller
than HfRhBi.

\section{Z$\text{r}$R$\text{h}$B$\text{i}$}

{\it Electronic Structure :} ZrRhBi is the most stable among the three HHA systems with the formation
energy of 725 meV/atom. It has a direct band gap of
1.021 eV, higher than the other two systems which may help in
minimizing the contribution of minority carrier contributions to the Seebeck coefficient,
there by increasing the ZT value. Figure \ref{dos+bs_ZrRhBi} shows the 
band structure and DOS for ZrRhBi. As seen from the band structure,
both the valence and conduction band edges are almost
flat near $\Gamma$ point. Also the DOS near E$_F$ shows a
sharp increase making the charge carrier heavier (higher effective
mass, helps in higher S value) which is a signature for it to be 
promising for both n-type and p-type TE applications.

{\it Thermoelectric properties:} ZrRhBi is found to have the highest values of TE figure of merit for both n-type
and p-type systems, which is mainly attributed to the nature
of bands (Figure \ref{dos+bs_ZrRhBi}). The
n$_c$ dependence of S, S$^{2}\sigma$, $\kappa$ and ZT at various
temperatures (300-1300 K) for both n-type and p-type system
are shown in Figure \ref{ZrRhBi}. For p-type ZrRhBi, at 1300 K, the highest
Seebeck coefficient (319.89 $\mu$VK$^{-1}$) among the three systems is obtained. The electrical
conductivity (8.7$\times$10$^4$ $\Omega^{-1}m^{-1}$) is very low as compared to
HfRhBi and ZrIrBi at optimal ($n_c$)$_{max}$ due to larger value of the band gap 
(making it difficult
for electrons to excite) and thus giving a smaller power factor $\approx$8.9
mWm$^{-1}$K$^{-2}$, compared to the other two systems.
Due to the large band gap, the electronic contribution to the thermal 
conductivity is also small because of the difficulty of thermally excited 
electron-hole pairs  to reach the conduction bands. In addition, the lattice
contribution to the thermal conductivity is also smaller compared to other
two systems due to the highly agitated phonons. Thus in spite of having the 
lowest electrical conductivity, ZrRhBi has the highest ZT value 
($\sim$ 0.43 at 1300 K) due to the extremely large S and lowest $\kappa$.

For n-type ZrRhBi, the Seebeck coefficients are slightly smaller than
the p-type (at high T) but much higher than the S values obtained
for n-type HfRhBi and ZrIrBi. Similar to p-type ZrRhBi, the electrical
conductivity is quite small, but large value of S makes the power
factor appreciable. ZT value in this case is equally large (0.45) as the 
p-type ZrRhBi.

\section{Discussion}

It is well known that electronic structure plays an important
role in determining the thermoelectric performance of a given material. In this
section, we will explain the reason for getting high ZT values in the three
candidate systems. Such plausible reasons can alone serve as a selection criteria
for choosing suitable parent materials, in general, with promising TE properties.

Figure \ref{dos+pdos} shows a closer view of the band structure and atom projected density of states very near to E$_F$ for the three systems. For HfRhBi and ZrIrBi, all the three constituent atoms have a slow rise in the number of states after E$_F$ indicating low effective mass values which in turn reduces the n-type Seebeck coefficients. However, the p-type Seebeck coefficient for these two materials are higher due to large DOS contribution (steep rise) of the atoms before E$_F$ (valence band edges).
In contrast, for ZrRhBi, it is interesting to note that the contribution of all the three constituent atoms to the conduction as well as valence band edge states show a steep rise in the DOS. This makes ZrRhBi to be a good TE material both as n- and p-type.

\begin{table*}[t!]
\begin{ruledtabular}
\begin{centering}
\begin{tabular}{|c|c|c|c|c|c|c|c|c|c|}

Systems & T & Doping type & $n_{c}$ & S & S$^{2}\sigma$ & $\sigma$ & $\kappa_{L}$ & $\kappa$ & (ZT)$_{max}$\tabularnewline
\hline 
\multirow{4}{*}{HfRhBi} & \multirow{2}{*}{1300} & n & 8.32 & 137.88 & 6.79 & 35.71 & \multirow{2}{*}{1.98} & 32.31 & 0.27\tabularnewline
\cline{3-7} \cline{9-10} 
 &  & p & 5.69 & 214.08 & 10.88 & 23.75 &  & 38.20 & 0.37\tabularnewline
\cline{2-10} 
 & \multirow{2}{*}{900} & n & 13.16 & 95.33 & 5.16 & 56.83 & \multirow{2}{*}{2.86} & 25.23 & 0.18\tabularnewline
\cline{3-7} \cline{9-10} 
 &  & p & 2.60 & 215.95 & 6.08 & 13.03 &  & 16.54 & 0.33\tabularnewline
\hline 
\multirow{4}{*}{ZrIrBi} & \multirow{2}{*}{1300} & n & 36.77 & 95.55 & 7.28 & 79.68 & \multirow{2}{*}{2.22} & 44.88 & 0.21\tabularnewline
\cline{3-7} \cline{9-10} 
 &  & p & 8.79 & 255.91 & 17.36 & 26.51 &  & 53.54 & 0.42\tabularnewline
\cline{2-10} 
 & \multirow{2}{*}{900} & n & 46.99 & 76.15 & 6.11 & 105.34 & \multirow{2}{*}{3.20} & 35.50 & 0.15\tabularnewline
\cline{3-7} \cline{9-10} 
 &  & p & 4.88 & 256.97 & 11.73 & 17.76 &  & 27.57 & 0.38\tabularnewline
\hline 
\multirow{4}{*}{ZrRhBi} & \multirow{2}{*}{1300} & n & 4.30 & 318.23 & 10.53 & 10.39 & \multirow{2}{*}{1.31} & 30.64 & 0.45\tabularnewline
\cline{3-7} \cline{9-10} 
 &  & p & 2.58 & 319.89 & 8.91 & 8.70 &  & 26.88 & 0.43\tabularnewline
\cline{2-10} 
 & \multirow{2}{*}{900} & n & 3.63 & 296.66 & 8.79 & 9.98 & \multirow{2}{*}{1.88} & 19.05 & 0.42\tabularnewline
\cline{3-7} \cline{9-10} 
 &  & p & 1.73 & 283.64 & 5.83 & 7.25 &  & 13.80 & 0.38\tabularnewline

\end{tabular}
\par\end{centering}
\caption{Highest ZT values for both n-type and p-type systems and corresponding
optimal TE parameters for HfRhBi, ZrIrBi and ZrRhBi at 1300 K and
900 K. T, $n_{c}$, S, S$^{2}\sigma$, $\sigma$, $\kappa_{L}$ and $\kappa$ are in the units of Kelvin, 10$^{20}$ cm$^{-3}$, $\mu$VK$^{-1}$, mWm$^{-1}$K$^{-2}$, 10$^4 $ $\Omega^{-1}$m$^{-1}$, mWm$^{-1}$K$^{-1}$ and mWm$^{-1}$K$^{-1}$ respectively.}
\label{final_data}
\end{ruledtabular}
\end{table*}

All the three systems discussed here are high temperature TE materials. At such high T, $\kappa_{L}$ 
contribution becomes very small compared to $\kappa_{e}$ (see Figure \ref{HfRhBi}, \ref{ZrIrBi}
 and \ref{ZrRhBi})  and hence can be neglected. This is because of the decrease in the mean free path 
($L_{ph}$) of the thermally agitated phonons which in turn reduces $\kappa_{L}$ ($\kappa_{L}$$\approx 
\frac{1}{3} v_{s}CL_{ph},$ where $v_{s}$ and $C$ are the velocity of sound through the material and
heat capacity, respectively). Thus ZT becomes,  $ZT_e = (S^{2}\sigma T)/\kappa_{e}$=$S^2/L$, using 
$\kappa_{e}=L\sigma T$. Therefore at high temperature, ZT essentially depends on the Seebeck coefficient
of the material. This decrease in mean free path of phonons is most prominent in ZrRhBi because of
  a large difference in the masses of the constituent atoms. This is also evident from the phonon 
dispersion curve (see supplement \cite{supl}), which reflect strong mass scattering and hence lowest lattice thermal conductivity.

With increase in temperature, electron-hole pairs are generated giving their individual contribution 
to the Seebeck coefficient. The total Seebeck coefficient can be expressed as, $ S= (S_{n}\sigma_{n}+S_{p}\sigma_{p} ) /(\sigma_{n}+\sigma_{p})$.
Since the Seebeck coefficient for electrons and holes have opposite
sign, the total S is usually lower than either of the individual contributions,
unless the direct band gap is large enough to minimize the minority
carrier contribution.

At moderate temperatures, the figure of merit calculated by only considering
electronic thermal conductivity (ZT$_{e}$) can be misleading
because in this temperature range, the $\kappa_{e}$ values are only somewhat 
larger or often even comparable to $\kappa_{L}$ values. The figure of merit can also
be expressed as,
$ZT=ZT_{e} (1+\frac{\kappa_{L}} {\kappa_{e}} )^{-1}$.
From the thermal conductivity curves (Figure \ref{HfRhBi}, \ref{ZrIrBi} and \ref{ZrRhBi})
it can be seen that at around 300-400 K, the $\kappa_{L}$ and $\kappa_{e}$ values 
are almost equal giving, $ZT=ZT_{e}/2$. Thus in the low temperature range, the role of
$\kappa_{L}$ is important and should not be ignored.

Since the Seebeck coefficient decreases with increase in carrier concentration
and increases with high effective mass while $\sigma$ increases with
small band gap and high mobility; suitable band topology, optimal
carrier concentration and band gap with high mobilities are desired
for promising TE applications. These factors were carefully analyzed 
before choosing the three systems discussed here. 
There are still open avenues to further enhance the thermoelectric efficiency 
of these materials via nanostructuring, doping with suitable elements etc.
The present work, however, sets an initial path to continue on these lines.

\section{Conclusion}
Over the years, there has been considerable efforts on enhancing the figure of merit (ZT) of
 thermoelectric materials by doping, surface, nano constructions etc. This enhancement, however,
very much depends on the ZT value of the parent ideal crystal. In the Half Heusler family, 
for example, the ZT value for the best parent materials (such as MNiSn (M=Ti, Zr, Hf), NbCoSb etc.) does not exceed beyond $\sim$0.3. One of the main motivation of the present paper is to
find new parent HHs systems, with already large ZT value. Taking a hint from the band structure
topology of a large combinatorial set of unreported HH compounds, we choose three systems, HfRhBi, ZrIrBi and 
ZrRhBi to perform a detailed electronic structure, electrical and thermal transport calculations for TE properties. 
All the three compounds are found to be chemically as well as mechanically strongly stable. They turn out to be efficient TE materials at high T. Table \ref{final_data} shows the optimal TE parameters (n$_c$, S, $\sigma$, S$^2 \sigma$ and $\kappa$) corresponding the highest ZT value for n- and p-type materials at two different temperatures, $1300$ and $900$ K. The figure of merit (ZT) for the three compounds lie in the range 0.21-0.45, with ZrRhBi having the largest value. Larger band gap ($\sim$1 eV) of ZrRhBi along with the occurrence of more flat bands in its band structure gives rise to higher S and hence the ZT value at high T. These compounds show an unusually high power factor ranging from 5.2 to 17.4 mWm$^{-1}$K$^{-2}$, responsible for high figure of merit. Although the highest possible power factor for the three systems HfRhBi, ZrIrBi and ZrRhBi are 15.49 mWm$^{-1}$K$^{-2}$, 22.57 mWm$^{-1}$K$^{-2}$ and 19.70 mWm$^{-1}$K$^{-2}$ respectively, however they do not correspond to the (ZT)$_{max}$ values (see Table \ref{final_data}), showing that choosing S$^2\sigma$ as deciding factor for optimal TE performance is highly misleading, as reported in many literature. The calculated maximum power factor, optimal n- and p-type carrier concentrations corresponding to maximum $ZT$ values for all the three systems certainly provide a guidance to the experimental work. Thermoelectric measurements on these new compounds are strongly recommended.

\section*{Acknowledgement}
AA acknowledges DST-SERB (SB/FTP/PS-153/2013) for funding to support this research. Enamullah (an institute post-doctoral fellow) acknowledges IIT Bombay for financial support.



\end{document}